\documentstyle[prl,aps,graphicx,floats]{revtex}
\begin{document}

\draft

\title{
Closing the Pseudogap by Zeeman Splitting in
Bi$_2$Sr$_2$CaCu$_2$O$_{8+y}$ at High Magnetic Fields}

\author{T. Shibauchi,$^{1,2}$ L. Krusin-Elbaum,$^{1}$
Ming Li,$^{3}$ M. P. Maley,$^{2}$ and P. H. Kes$^{3}$}

\address{$^{1}$IBM T.~J. Watson
Research Center, Yorktown Heights, New York 10598, USA}
\address{$^{2}$Superconductivity Technology Center, Los Alamos National Laboratory,
Los Alamos, New Mexico 87545, USA}
\address{$^{3}$Kamerlingh Onnes Laboratory, Leiden University,
2300 RA Leiden, The Netherlands }

\date{Received 27 November 2000} \wideabs{

\maketitle

\begin{abstract}

Interlayer tunneling resistivity is used to probe the low-energy
density-of-states (DOS) depletion due to the pseudogap in the
normal state of Bi$_2$Sr$_2$CaCu$_2$O$_{8+y}$.  Measurements up
to 60~T reveal that a field that restores DOS to its ungapped
state shows strikingly different temperature and doping
dependencies from the characteristic fields of the
superconducting state. The pseudogap closing field and the
pseudogap temperature $T^{\star}$ evaluated independently are
related through a simple Zeeman energy scaling. These findings
indicate a predominant role of spins over the orbital effects in
the formation of the pseudogap.

\end{abstract}
}

\narrowtext


A central unresolved issue of high temperature superconductivity
is the connection of normal state correlations, referred to as
the pseudogap \cite{Timusk,Norman,Watanabe2,Suzuki,Krasnov}, to
the origins of high-$T_c$. At the heart of the debate
\cite{Xu,Mitrovic,Gorny,Zheng,Zheng2,Krasnov2} is whether the
pseudogap, which manifests itself as a depletion of the
quasiparticle density of states (DOS) below a characteristic
temperature $T^\star$, originates from spin or charge degrees of
freedom and, in particular, whether it derives from some
precursor of Cooper pairing \cite{Emery} that acquires the
superconducting coherence at $T_c$. Energies of the order of the
pseudogap have been accessed with elevated temperatures, with
applied voltage in tunneling measurements, and with infrared
frequencies in optical spectra \cite{Timusk}. But little is known
about the effect of magnetic field. The magnetic field response
may be unique: {\it e.g.}, in the case of the superconducting
state the upper critical field $H_{c2}$ is determined by the
superconducting coherence length, and not directly by the
superconducting gap, since magnetic field strongly couples to the
orbital motion of Cooper pairs.

Current knowledge about the field dependence of the pseudogap
derived from spectroscopic measurements is partly limited by the
available dc field range
\cite{Mitrovic,Gorny,Zheng,Zheng2,Krasnov2}. More importantly,
there is no systematic doping dependence in a single family of
cuprates. Even in optimally doped YBa$_2$Cu$_3$O$_{7-\delta}$
alone, based on NMR relaxation rate measurements below 27.3~T,
the pseudogap was claimed to decrease \cite{Mitrovic} or be
independent of magnetic field \cite{Gorny}. In the underdoped
YBa$_2$Cu$_4$O$_8$ no field effect on $T^\star$ was reported up
to 23.2~T \cite{Zheng}, while a recent NMR study indicated a
measurable field dependence in slightly overdoped
TlSr$_2$CaCu$_2$O$_{6.8}$ \cite{Zheng2}. In this paper, we report
the interlayer ($c$-axis) resistivity $\rho_c$ measurements in
fields up to 60~T in Bi$_2$Sr$_2$CaCu$_2$O$_{8+y}$ (BSCCO)
crystals in a wide range of doping, from which we make a first
systematic evaluation of the pseudogap closing field $H_{pg}$
that restores DOS to its ungapped state. Our results indicate a
pronounced difference between field-temperature ($H$-$T$)
diagrams of the pseudogap and the superconducting states and a
simple Zeeman scaling between $H_{pg}(0)$ and $T^\star$.

Among various techniques that quantify DOS, the $\rho_c$ measurements are uniquely suited for
exploring the highest magnetic field range available only in a pulsed mode. In highly anisotropic
materials such as BSCCO where interlayer coupling between CuO$_2$ layers is sufficiently weak, the
$c$-axis transport directly measures Cooper pair or quasiparticle tunneling in both normal and
superconducting states \cite{Morozov}, providing {\em bulk} information about the quasiparticle
DOS at the Fermi energy. Thus, $\rho_c$ should be particularly sensitive to the onset of the
pseudogap formation, since the DOS depletion is largest at the Fermi energy. Moreover, $\rho_c$ is
controlled by the ($\pi$, 0) points (`hot spots') on the anisotropic Fermi surface
\cite{Ioffe,Valla}, where the pseudogap first opens up \cite{Norman}. (This is in contrast to the
in-plane resistivity $\rho_{ab}$ mainly determined by carriers with momenta along the ($\pi$,
$\pi$) directions\cite{Ioffe}.)

Our task here is to map $H$-$T$ diagram of the pseudogap state.
To elucidate the field dependence of the pseudogap over a wide
doping range, we carefully adjusted hole concentration $p$
spanning both underdoped and overdoped regimes in BSCCO crystals
grown by the floating-zone method \cite{Mingli}. The doping level
was controlled by annealing in O$_2$ or N$_2$ at the appropriate
pressures \cite{Watanabe2}. $\rho_c(H)$ was measured using a 33~T
dc magnet \cite{Brandt} and a 60~T long pulse (LP) system at the
National High Magnetic Field Laboratory (NHMFL) \cite{heating}.

In slightly underdoped BSCCO at temperatures below $T_c$, the
field dependence of $\rho_c$ exhibits a peak that we have
previously demonstrated to arise from a competition between two
tunneling conduction channels: of Cooper pairs (at low fields)
and quasiparticles (mainly at high fields) \cite{Morozov}. The
peak position marks the field (in the superconducting state)
where the quasiparticle contribution overtakes the Cooper pair
tunneling current. The doping dependence of $\rho_c(H)$ in Fig.~1
clearly shows that the peak field $H_{sc}$ in the highly
underdoped crystal, where interlayer (Josephson) coupling is the
weakest, is most easily suppressed. Magnetoresistance (MR) above
$H_{sc}$ is negative and remains so above $T_c$, as has been seen
at lower fields \cite{Yan}. An important difference between the
underdoped and overdoped regimes is in the slope of the negative
MR. A gentle slope in the underdoped regime turns steeper in the
overdoped regime and, as shown in Fig.~1(c), at the highest
fields the low-temperature $\rho_c(H)$ rapidly approaches the
normal-state value.

In the overdoped samples, negative MR eventually disappears. This
occurs at the same temperature at which the zero-field
$\rho_c(T)$ develops a characteristic upturn from the $T$-linear
dependence of the metallic state [Fig.~2(a)]. This temperature --
at which a gap-like feature also appears in the static
susceptibility \cite{Watanabe2} and in the tunneling spectra
\cite{Suzuki} of BSCCO -- is identified as the pseudogap
temperature $T^\star$. In the pseudogap state below $T^\star$,
the negative MR is naturally understood by the suppression of the
pseudogap by magnetic field \cite{note}. In our most overdoped
crystal with $T_c = 67$~K, a magnetic field of $\sim 60$~T
downshifts the $\rho_c(T)$ upturn and the associated $T^\star$ by
about 20~K [Fig. 2(b)]. In other words, at this doping level, the
60~T field at $\sim 100$~K closes the pseudogap. To track the
pseudogap closing field at lower temperatures, we consider the
excess resistivity $\Delta\rho_c$ due to the DOS depletion. It is
known from the intrinsic tunneling spectroscopy measurements
\cite{Krasnov} that the $T$-linear dependence of the metallic
state persists below $T^\star$ for bias voltages sufficiently
above the pseudogap voltage. Subtracting this metallic
contribution gives $\Delta\rho_c$ \cite{Deltarhoc}. The field at
which $\Delta\rho_c$ vanishes is the pseudogap closing field
$H_{pg}(T)$. To obtain $H_{pg}(T)$, we first note that the
$c$-axis MR in the pseudogap state was recently shown to follow a
power-law field dependence up to 60~T \cite{Morozov}. A fit to
the power-law field dependence of $\Delta\rho_c(H)$ at different
temperatures [inset in Fig.~2(b)] allows us to evaluate
$H_{pg}(T)$ beyond 60~T. This evaluation is insensitive to the
detailed functional form of the fit, as can be inferred from
Fig.~1(c). We tried other extrapolation fits (e.g. polynomial)
and they gave the same values of $H_{pg}(T)$ within the error
bars in Fig.~3(a).

The entire $H$-$T$ diagram of the pseudogap in the overdoped
crystal is shown in Fig.~3. At low temperatures $H_{pg}$ is
essentially flat with the limiting value of $\sim 90$~T. This is
in marked contrast with the characteristic fields of the
superconducting state: the peak field $H_{sc}(T)$ and the
irreversibility field $H_{irr}(T)$. At low temperatures
$H_{sc}(T)$ grows exponentially and points to the
zero-temperature value of $\sim 100$~T, nearly independent of
doping [Fig.~3(b)]. This difference, consistent with recent NMR
\cite{Gorny,Zheng} and intrinsic tunneling measurements
\cite{Krasnov2}, may indicate different origins of the pseudo-
and superconducting gaps.

In the overdoped crystals the low-temperature $H_{pg}$ and the
zero-field $T^\star$ can be obtained independently and the
comparison between the two (Fig.~4) leads to a strikingly simple
conclusion. The pseudogap closing field scales with $T^\star$ as
$g\mu_BH_{pg} \approx k_BT^\star$. Here $g$-factor $g=2.0$,
$\mu_B$ is the Bohr magneton, and $k_B$ is the Boltzmann
constant. This immediately suggests that magnetic field couples
to the pseudogap by the Zeeman energy of the spin degrees of
freedom. In the underdoped regime, the appreciable error bars in
$H_{pg}$ reflect the fact that the extrapolation extends
considerably beyond the maximum laboratory field range of 60~T
[inset of Fig.~1(c)]. However, the estimate gives a consistent
and physically sensible picture, since in the underdoped regime
we find that below 150~K $H_{pg}(T)$ is also flat and $H_{pg}(p)$
is a smooth continuation from the overdoped side. The observed
general trend of the high-field slope of $\rho_c(H)$ as a
function of doping is unmistakable, and the Zeeman energy scale
of $H_{pg}$ is in good agreement with the reported energy scale
of the pseudogap and $T^\star$ \cite{Timusk,Miyakawa} (see the
shaded band in Fig.~4). Thus, we surmise that the Zeeman scaling
found in the overdoped samples holds in the entire doping range.

In contrast to $H_{pg}(p)$, the doping dependence of the peak
field $H_{sc}$ is weak and roughly follows a parabolic dependence
similar to $T_c(p)$. Note that $H_{sc}$ does not represent an
upper critical field $H_{c2}$. The boundary at $H_{c2}$ in
high-$T_c$ superconductors is a fuzzy crossover difficult to
estimate \cite{Blatter}, but should be higher than $H_{sc}$
\cite{Morozov}. If strong phase fluctuations associated with the
precursor superconductivity are responsible for the pseudogap
state \cite{Emery}, one would expect $H_{pg} \gg H_{c2} \gtrsim
H_{sc}$ in the zero-temperature limit since quantum phase
fluctuations should also be significant. In the underdoped
regime, the difference between $H_{pg}$ and $H_{sc}$ in the low
temperature limit is huge, which in this scenario can be
attributed to quantum fluctuations. Surprisingly, in the
overdoped regime, while the pseudogap region in the $H$-$T$
diagram is still nontrivially large, the low temperature values
of $H_{pg}$ and $H_{sc}$ are nearly the same, which may raise
questions about large quantum fluctuations.

Our finding that Zeeman splitting closes the pseudogap implies
that the triplet spin excitation at high fields overcomes the
singlet pair correlations responsible for the gap in the spin
spectrum, and that the orbital contribution is very small. In
preformed pair scenarios, our results would require pairing
correlations on relatively short length scales with negligible
orbital motion of pairs. This may be satisfied in a class of
models, where charges (holes) self-organize into micro-stripes
below $T^\star$ \cite{Emery2,Tsuei}. The mechanism of pairing is
the generation of the `spin-gap' in spatially confined
Mott-insulating regions with local antiferromagnetic correlations
in the proximity of the metallic stripes \cite{Emery2}. However,
any future theoretical input must reconcile such pairing with a
lack of significant quantum fluctuations in the overdoped regime.

A spin-gap in a doped Mott insulator also appears in
resonating-valence-bond theory, where the spin and charge degrees
in the CuO$_2$ plane are separated into `spinons' and `holons'
\cite{Anderson}. Studies based on this idea \cite{Lee} derive a
doping-dependent spin-gap temperature evolving from zero on the
overdoped side to a finite value prescribed by the
antiferromagnetic exchange $J$ ($\sim 1000$~K) as $p\rightarrow
0$. This spin-gap temperature corresponds to the formation of
spinon singlet pairs. A gap in the spin excitation spectrum can
be seen in the $c$-axis tunneling spectra, since during the
interplane tunneling process spinons and holons recombine into
conventional carriers with charge and spin. Our empirical linear
scaling of $H_{pg}(p)$ and $T^\star(p)$ gives an energy scale
$\sim 930$~K in the $p\rightarrow 0$ limit, of the order of $J$.

Our results up to 60~T point to a predominant role of spins in
the formation of the pseudogap consistent with models based on a
doped Mott insulator \cite{Emery2,Tsuei,Anderson,Lee}. An
interesting issue concerns the existence of a quantum critical
point at which the pseudogap temperature goes to zero
\cite{Chakravarty}. This has been argued to occur at a critical
doping $p \approx 0.19$ \cite{Tallon}. However, our most
overdoped crystal with $p \approx 0.22$ has the pseudogap still
unmistakably prominent \cite{Suzuki2}, likely reflecting higher
sensitivity of the interlayer tunneling at ($\pi,0$) points on
the Fermi surface \cite{Ioffe,Valla}, where a spectral weight
depletion onsets at a higher temperature \cite{Norman}.

We thank N. Morozov, F. F. Balakirev, J. Betts, C. H. Mielke, and
B. L. Brandt for technical assistance, and L. N. Bulaevskii, N.
Nagaosa, V. B. Geshkenbein, L. B. Ioffe, C. C. Tsuei, and D. M.
Newns for helpful discussions. This work was supported in part by
NSF through NHMFL by the Contract No. AL99424-A009. Measurements
were performed at NHMFL, which is supported by the NSF
Cooperative Agreement No. DMR-9527035.


\normalsize
\begin{figure}[tb]
\begin{center}
\includegraphics[scale=0.37]{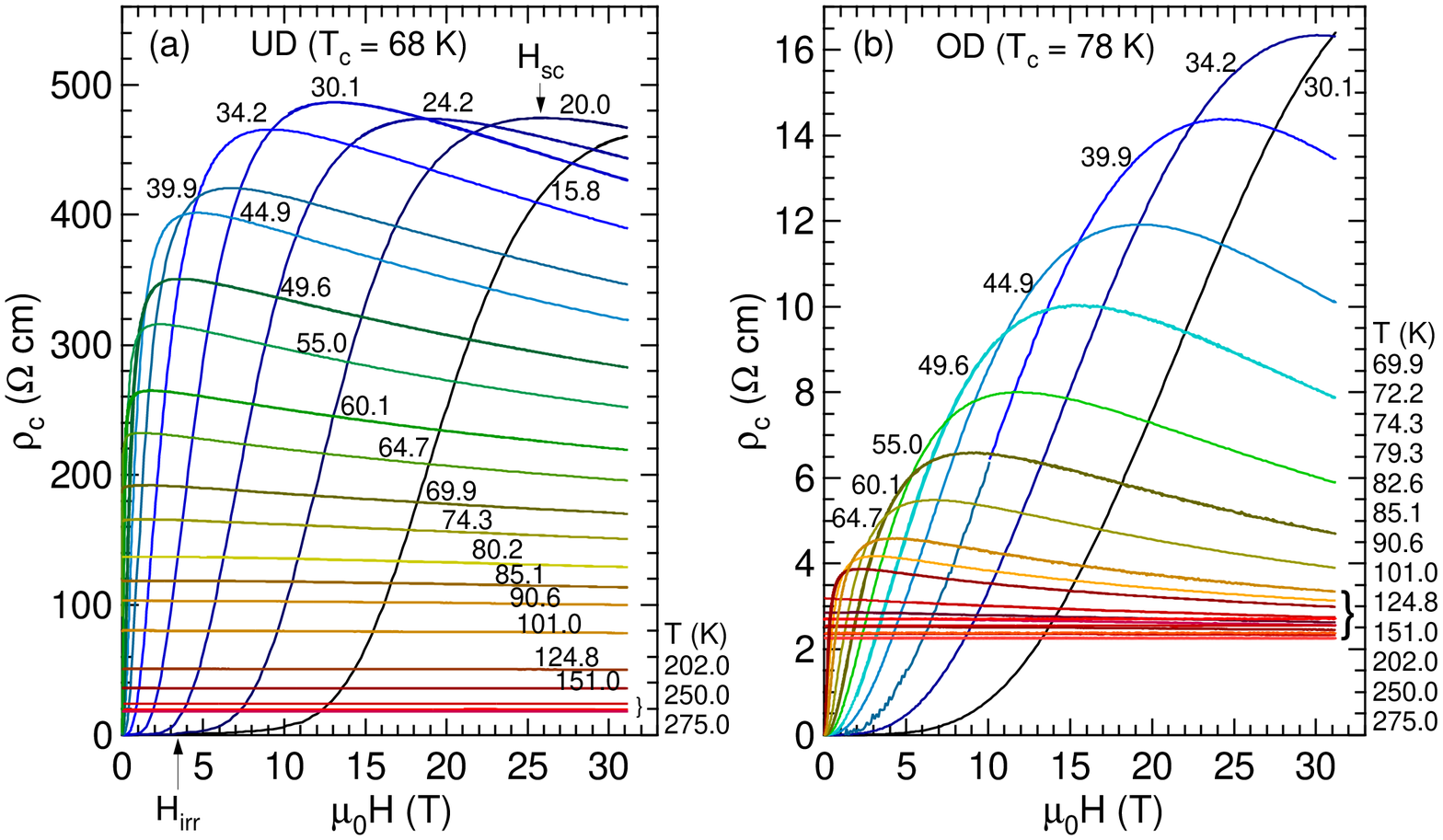}
\includegraphics[scale=0.37]{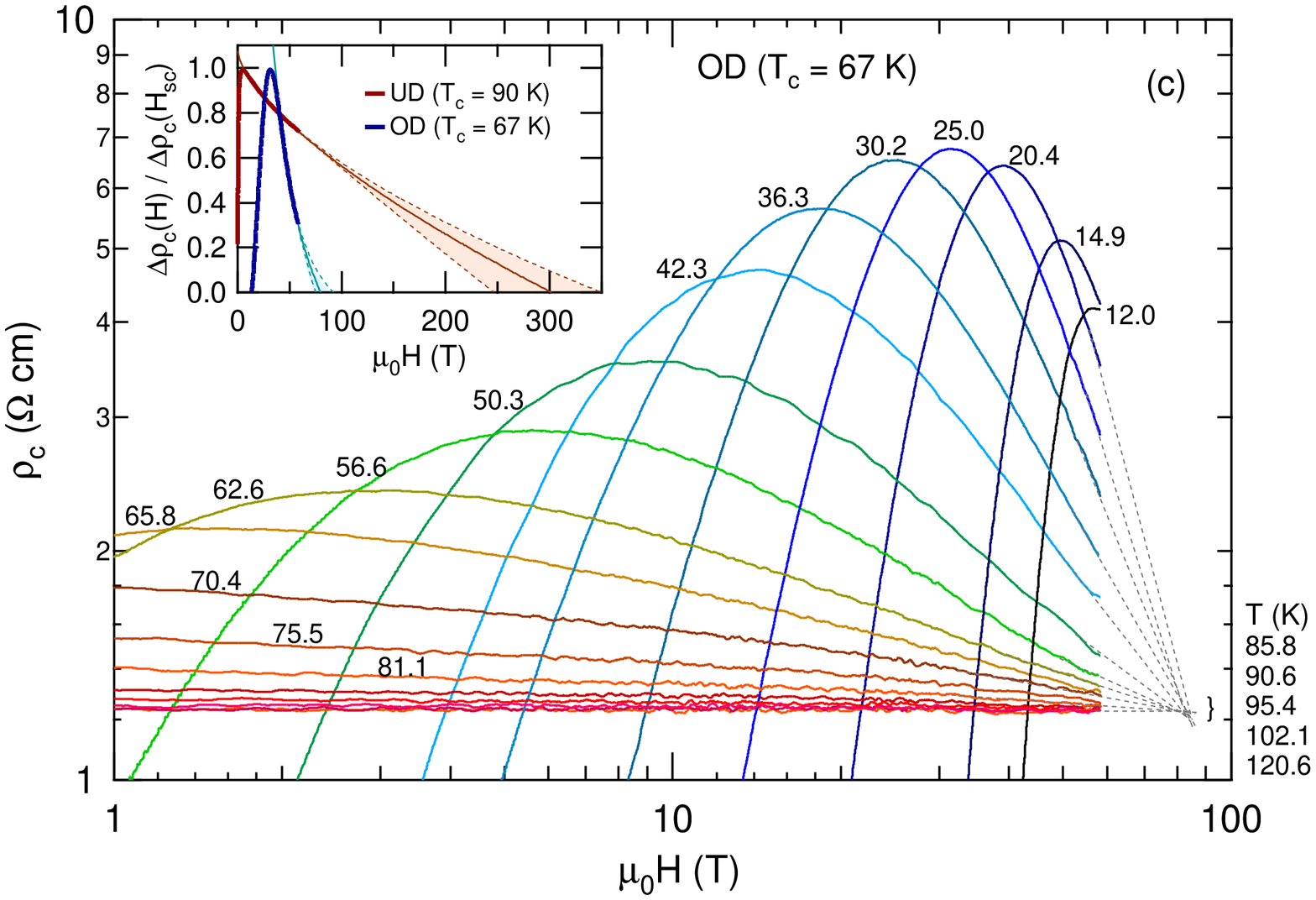}
\end{center}
\vspace{5mm}
\caption{$c$-axis resistivity $\rho_c$ (labeled by temperatures)
vs magnetic field $H (\parallel c)$ in an underdoped (UD) BSCCO
crystal (a) and two overdoped (OD) crystals (b), (c). In the
superconducting state, $\rho_c(H)$ becomes finite above the
irreversibility field $H_{irr}$ and exhibits a peak at $H_{sc}$.
The core feature in $\rho_c(H)$ that changes with doping is the
slope of the high-field negative MR. Dotted lines in (c) are
guide to the eye pointing to the limiting value ($<100$~T) of
$H_{pg}(T)$. Inset: Excess resistivity due to the pseudogap
$\Delta\rho_c$ (see Fig.~2) as a function of field for two
samples at $T \sim 0.2 T^\star$. Thin lines are power-law fits
and the shades indicate the uncertainties estimated from the
leeway in the fitting parameters. }
\end{figure}

\begin{figure}[tb]
\begin{center}
\includegraphics[scale=0.45]{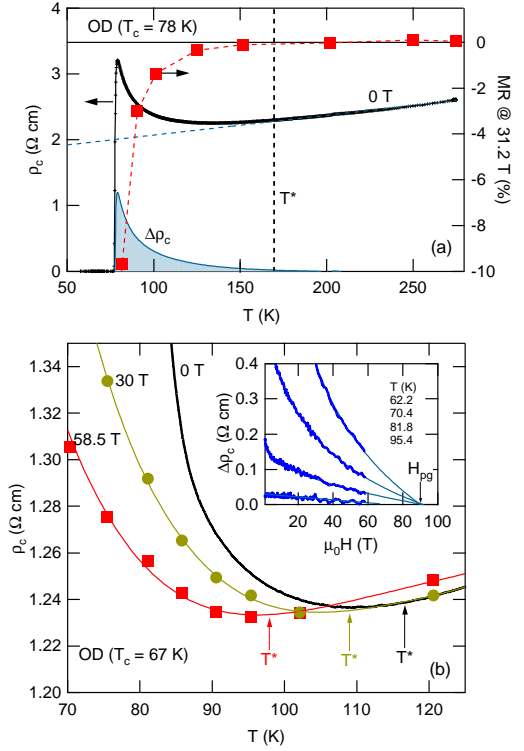}
\end{center}
\vspace{5mm}
\caption{Determination of the pseudogap temperature $T^\star$ and
the pseudogap closing field $H_{pg}$ from $\rho_c(T, H)$ in
overdoped BSCCO. (a) $\rho_c(T)$ deviates from metallic
$T$-linear dependence at the same temperature where negative MR
disappears, identified as pseudogap temperature $T^\star$. (b) In
our most overdoped BSCCO with $T_c=67$~K, $T^\star$ is shifted by
$\sim 20$~K by a 58.5~T field. Inset: The excess quasiparticle
resistivity $\Delta\rho_c(H)$ (above $H_{sc}$) is fitted to a
power-law field dependence
$[\Delta\rho_c(H)-\Delta\rho_c(0)]\propto H^{\alpha}$.
}
\end{figure}

\begin{figure}[tb]
\begin{center}
\includegraphics[scale=0.4]{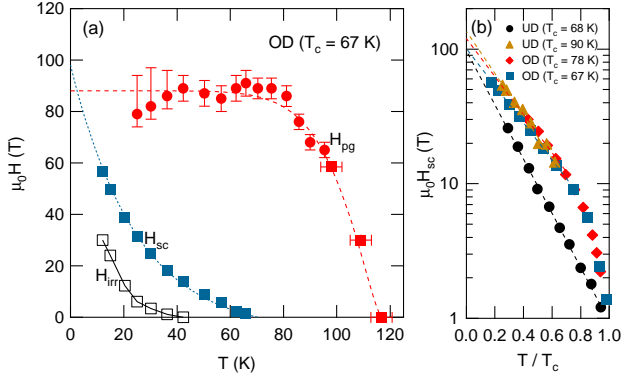}
\end{center}
\vspace{5mm}
\caption{$H$-$T$ diagram showing the pseudogap closing field
$H_{pg}(T)$, the peak field $H_{sc}(T)$, and the irreversibility
field $H_{irr}(T)$ in the overdoped BSCCO. (a) Up to 60~T,
$H_{pg}(T)$ is directly determined from the down-shifting upturn
of $\rho_c(T)$ (red squares). At lower temperatures, $H_{pg}(T)$
is obtained by extrapolating $\Delta\rho_c(H)$ to zero [inset in
Fig.~2(b)]. The two procedures consistently produce a seamless
$H_{pg}(T)$ within the error bars. The starkly different
temperature dependencies of $H_{pg}(T)$ and $H_{sc}(T) \le
H_{c2}$ here extrapolate to roughly the same zero-temperature
value. The usual estimate $H_{c2}(0) = 0.7 (\partial H_{c2}
/\partial T)|_{T_c}T_c$ with an initial slope of $\sim 2$~T/K 
[13] gives $H_{c2}(0) \approx 94$~T, very close to the value of 
$H_{sc}(0)$. (b) At low temperatures, $H_{sc}(T)$ grows nearly 
exponentially and, in the $T \rightarrow 0$ limit, is only weakly 
dependent on doping. }
\end{figure}

\begin{figure}[tb]
\begin{center}
\includegraphics[scale=0.5]{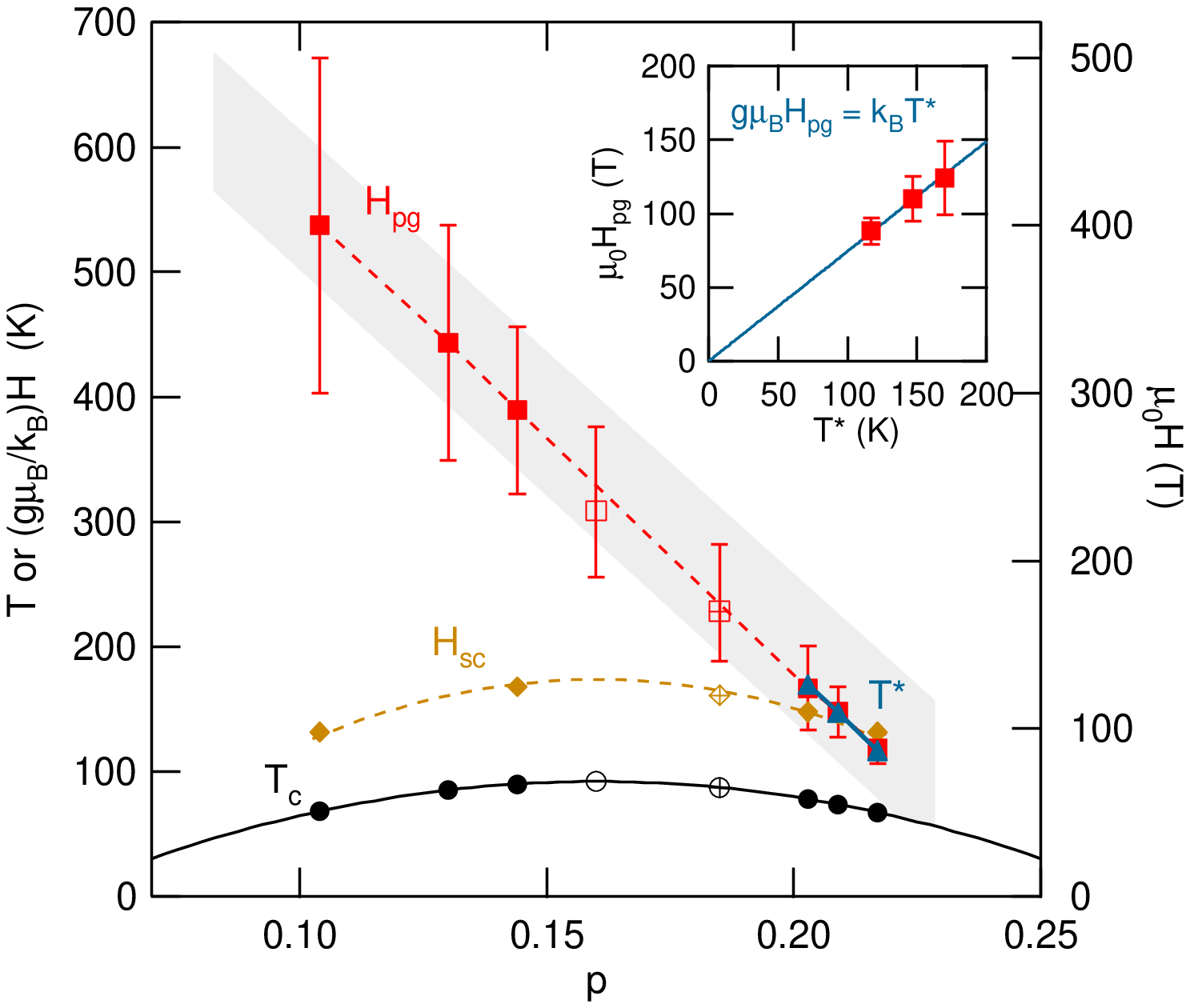}
\end{center}
\vspace{5mm}
\caption{Doping dependencies of low-temperature $H_{pg}$, 
$H_{sc}(T\rightarrow 0)$, $T^\star$, and $T_c$. The hole 
concentration $p$ was obtained from the empirical formula 
$T_c/T_c^{max} = 1-82.6(p-0.16)^2$ [29] with $T_c^{max}=92$~K. 
The right-hand-side field scale directly translates onto the 
Zeeman energy scale on the left-hand-side as $(g\mu_B/k_B)H$. 
$H_{pg}$ (red squares) and $T^\star$ (blue triangles), obtained 
separately in the same crystals in the overdoped regimes, give a 
scaling $g\mu_BH_{pg} \approx k_BT^\star$ with $g=2.0$ (inset). 
Open and crossed symbols are from our analysis of $\rho_c(H, T)$ 
in Refs.~[31] and [32], respectively. The shaded band covers 
$T^\star(p)$ in cuprates [22] taken from Fig.~26 of Ref.~[1].
}
\end{figure}


\begin{thebibliography}{9}
\bibitem{note} The pseudogap persists below $T_c(H=0)$
(see Fig.~3(a) and also Ref.~\protect{\onlinecite{Krasnov}}). In
the vicinity of $H_{c2}$ estimated from magnetic measurements,
$\rho_c(H)$ shows no distinct anomaly
\protect{\onlinecite{Morozov}} (negative MR extends beyond
$H_{c2}$), but the value of $\rho_c$ remains much higher than the
expected value in the ungapped state, indicating the dominant
contribution from the pseudogap to the low-energy DOS at high
fields. At the lowest temperatures the regime above $H_{c2}$
cannot be accessed.
\bibitem{Deltarhoc} In the underdoped regime, the metallic
contribution is negligible compared to $\Delta\rho_c(H=0)$.
\end{thebibliography}
\end{document}